\documentclass[final,3p,times,twocolumn]{elsarticle}

\usepackage{amsthm}
\usepackage{amssymb}
\usepackage{amsmath}
\usepackage{color}
\usepackage{subfigure}
\usepackage{url}

\usepackage[switch]{lineno}

\usepackage{diagbox}
\usepackage[colorlinks,linkcolor=blue,anchorcolor=blue,citecolor=blue]{hyperref}
\biboptions{sort&compress}
\usepackage{threeparttable}

\journal{NIM A}

\begin{document}

\begin{frontmatter}



\title{Characterization of 4H-SiC Low Gain Avalanche Detectors (LGADs)}

\author[label1]{Tao Yang}
\author[label2,label3]{Ben Sekely}
\author[label2,label3]{Yashas Satapathy}\author[label4]{Greg Allion}
\author[label4]{Philip Barletta}
\author[label1]{Carl Haber \corref{cor}}
\ead{chhaber@lbl.gov}
\author[label1]{Steve Holland}
\author[label2]{John F. Muth \corref{cor}}
\ead{muth@ncsu.edu}
\author[label3]{Spyridon Pavlidis}
\author[label5]{Stefania Stucci}

\affiliation[label1]{organization={Physics Division, Lawrence Berkeley National Laboratory},
            addressline={1 Cyclotron Road},
            city={Berkeley},
            postcode={94720},
            state={CA},
            country={U.S.}}
\affiliation[label2]{organization={Department of Materials Science and Engineering, North Carolina State University},
            addressline={Engineering Bldg I, 911 Partners Way},
            city={Raleigh},
            postcode={27606},
            state={NC},
            country={U.S.}}
\affiliation[label3]{organization={Department of Electrical and Computer Engineering, North Carolina State University},
            addressline={2410 Campus Shore Dr},
            city={Raleigh},
            postcode={27606},
            state={NC},
            country={U.S.}}
\affiliation[label4]{organization={NCSU Nanofabrication Facility, North Carolina State University},
            addressline={2410 Campus Shore Dr},
            city={Raleigh},
            postcode={27606},
            state={NC},
            country={U.S.}}
\affiliation[label5]{organization={Brookhaven National Laboratory},
            addressline={98 Rochester St},
            city={Upton},
            postcode={11973},
            state={NY},
            country={U.S.}}

\cortext[cor]{Corresponding authors}

\begin{abstract}

4H-SiC low gain avalanche detectors (LGADs) have been fabricated and characterized. The devices employ a circular mesa design with low-resistivity contacts and an SiO$_2$ passivation layer. The I-V and C-V characteristics of the 4H-SiC LGADs are compared with complementary 4H-SiC PiN diodes to confirm a high breakdown voltage and low leakage current. Both LGADs and PiN diodes were irradiated with $\alpha$ particles from a $^{210}_{84}\rm{Po}$ source. The charge collected by each device was compared, and it was observed that low-gain charge carrier multiplication is achieved in the 4H-SiC LGAD.

\end{abstract}


% \begin{highlights}
% \end{highlights}


\end{frontmatter}

% \linenumbers


\begin{figure*}[htb]
    \centering
    \subfigure[]{ \label{fig:LGAD_3D_Cross_Section}
        \includegraphics[scale=0.45]{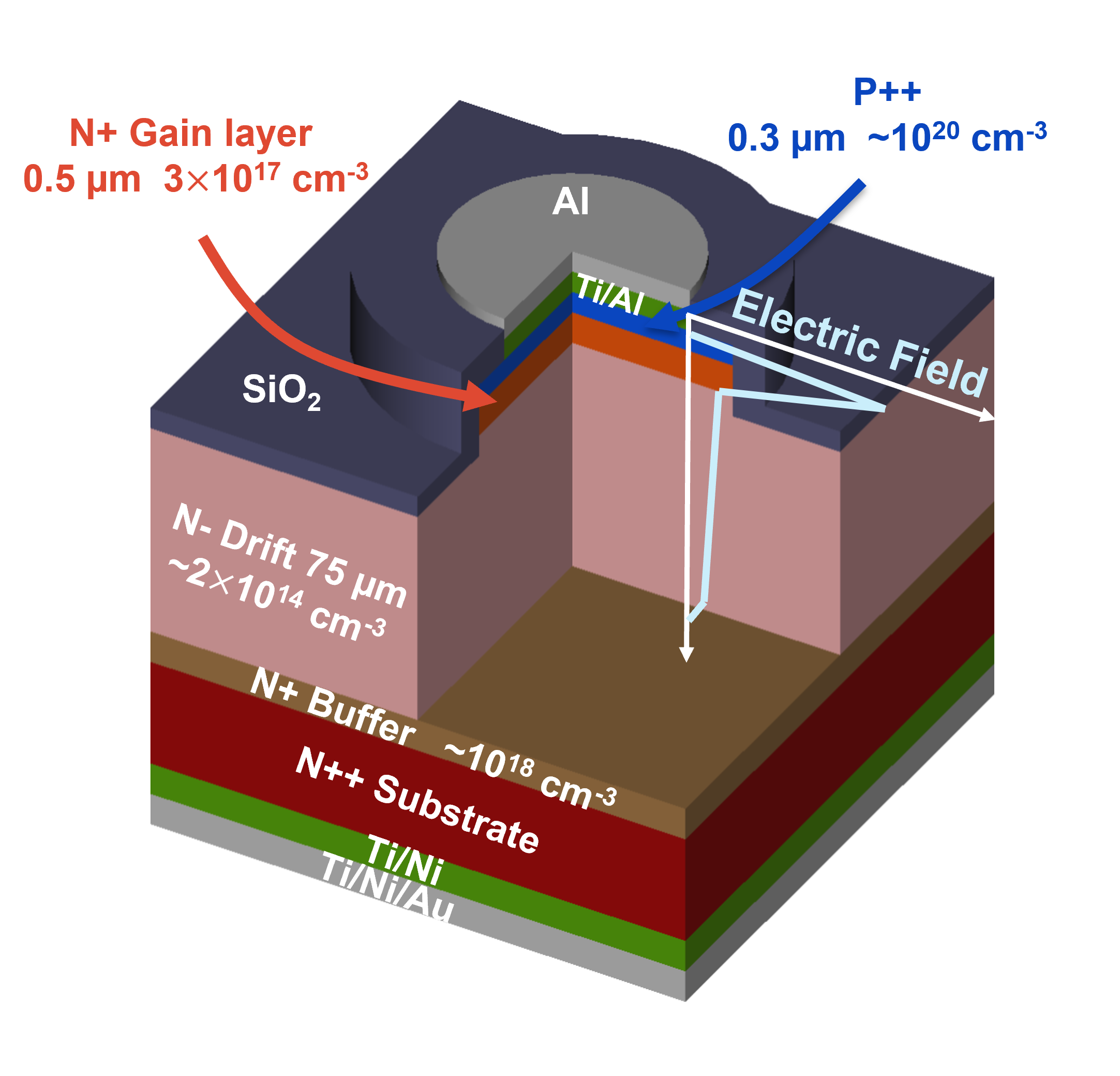}}  
    \subfigure[]{ \label{fig:PIN_3D_Cross_Section}
        \includegraphics[scale=0.45]{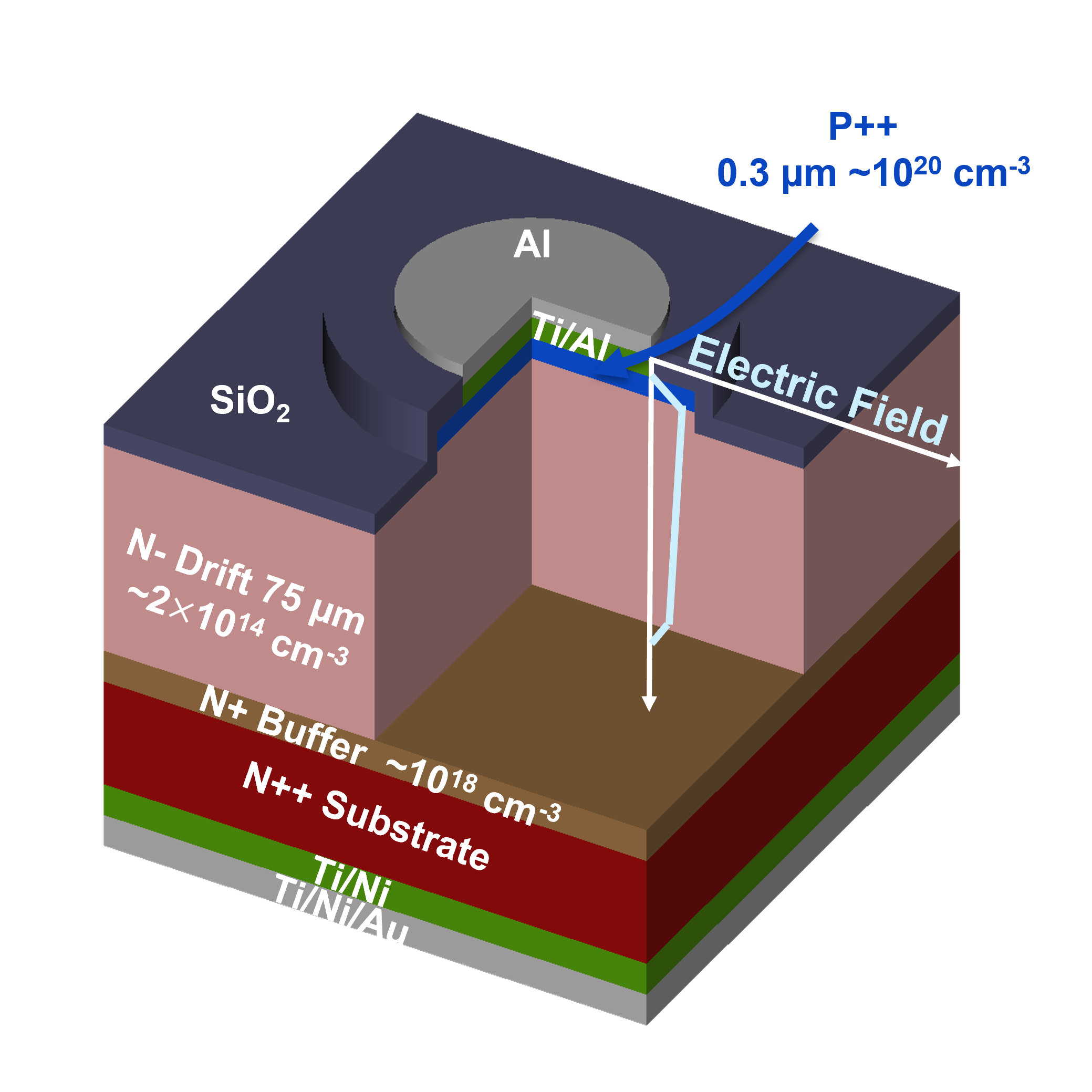}}
    \caption{Epi-stack of the circular (a)~4H-SiC LGAD and (b)~4H-SiC PiN diode. The electric field for the devices are shown.}
\end{figure*}

\section{Introduction} \label{sec:introduction}
Over the past decade, silicon (Si) low gain avalanche detectors (LGADs) have demonstrated excellent time resolution ($<$50~ps) and have been successfully fabricated by numerous institutions worldwide \cite{CNM_LGAD,HPK_LGAD,FBK_LGAD,BNL_LGAD,KeweiWU_LGAD,YunyunFAN_NDL}. Si LGADs have been designated for use in the ATLAS \cite{ATLAS_HGTD} and CMS \cite{CMS_ETL} detectors at the Large Hadron Collider (LHC), and they are also planned for future use in the Electron-Ion Collider (EIC) experiments \cite{EIC_Concept}. Si LGADs, irradiated to fluences expected at hadron colliders, are seen to exhibit increased leakage current and gain degradation due to acceptor removal effects \cite{LGAD_RAD_I_G,Ar_Model}. To address radiation damage effects, a common mitigation is to cool the sensors to -30 $^{\circ}$C. This increases the operational and manufacturing costs of the detectors and adds to the material budget of the detector.

Silicon carbide (SiC) is a wide bandgap semiconductor. Among its many polytpes, 4H-SiC is the most commonly used in commercial (e.g., power electronics) applications due to its large critical electric field, high mobility and rapidly expanding native substrate availability. Compared to Si, the larger bandgap of 4H-SiC also results in lower leakage current and higher theoretical radiation hardness. In addition, 4H-SiC has a higher carrier saturation drift velocity, higher atomic displacement threshold energy, higher thermal conductivity, and lower leakage current after irradiation \cite{SiC_Rad}. Therefore, 4H-SiC detectors have the potential to outperform their Si counterparts for fast charged particle detection in collider detectors. Indeed, previous work has verified that 4H-SiC detectors exhibit good timing resolution ($\sim$100~ps) \cite{SiC_Alpha_Time,Tao_NJU_PIN_Time}. Therefore, combining the excellent timing performance of LGADs with the advantages of SiC material may enable 4H-SiC LGADs to operate effectively in high-radiation environments while providing precise time measurements.

The concept of the 4H-SiC LGAD was first proposed in \cite{First_SiC_LGAD_Report}, and in the following years, several simulation designs and prototype devices were reported \cite{NJU_SiC_LGAD_Report_2,IHEP_SiC_LGAD_Report,Tao_Design_SiC_LGAD}. These prototypes described or used the LGAD structure but did not verify through comparative experiments whether low-gain carrier multiplication was achieved. Nonetheless, these results provided important data references for the realization of a true 4H-SiC LGAD.

In this work, we designed wafers with 4H-SiC epitaxial (epi-) stacks having specific doping concentrations to produce the typical electric field distribution of LGADs. For comparison, we also acquired wafers without the gain layer, resulting in a typical PiN diode electric field distribution. By comparing the measurement results of 4H-SiC LGADs and 4H-SiC PiN diodes fabricated from these wafers, we demonstrated low-gain carrier multiplication in the 4H-SiC LGADs. The collected charge in the 4H-SiC LGADs significantly increased with higher bias voltage, indicating that the gain factor increases with the electric field in the gain layer.

\section{Device Design and Fabrication} \label{sec:device_fabrication}

\begin{figure}[htb] 
    \centering
    \includegraphics[scale=0.55]{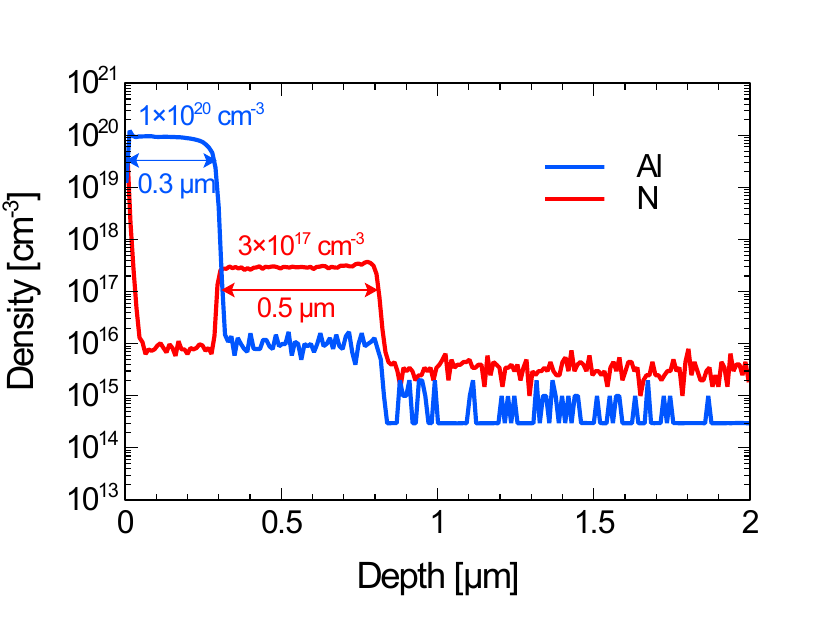}
    \caption{Doping concentration obtained by secondary ion mass spectrometry: blue points are aluminum and red points are nitrogen.}
    \label{fig:W2_SIMS}
\end{figure}

\begin{figure}[htb] 
    \centering
    \includegraphics[scale=0.55]{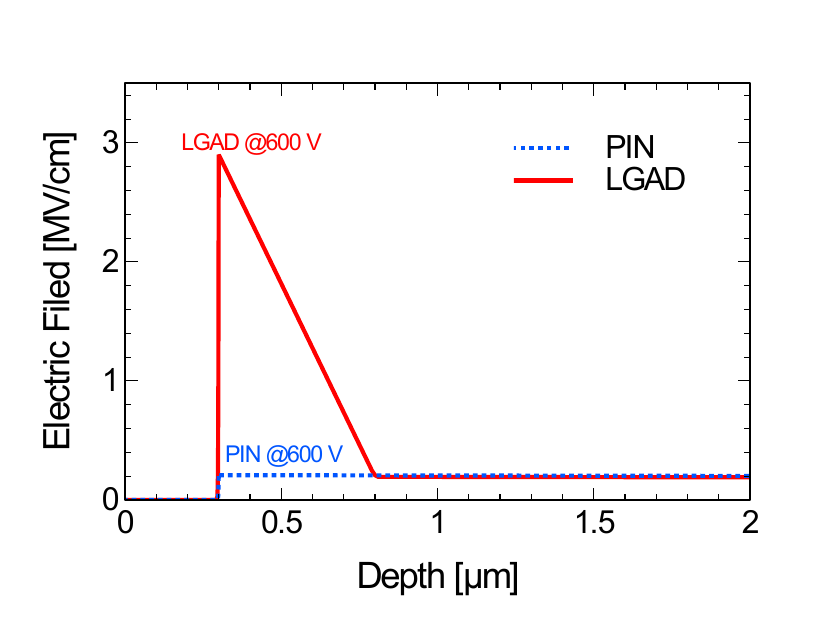}
    \caption{Simulated electric field distribution of the LGAD (red) and PiN diode (blue) at the 600~V reverse bias voltage.}
    \label{fig:ElectricField}
\end{figure}

As shown in \figurename~\ref{fig:LGAD_3D_Cross_Section}, the epi-structure of the LGAD includes a p$^{++}$ contact layer (0.3 $\mu$m; 1$\times$10$^{20}$ cm$^{-3}$), n$^+$ gain layer (0.5 $\mu$m; 3$\times$10$^{17}$ cm$^{-3}$), n$^-$ drift layer (75 $\mu$m; 2$\times$10$^{14}$ cm$^{-3}$), and n$^+$ buffer layer ($\leq$ 1 $\mu$m; $\geq$5$\times$10$^{18}$ cm$^{-3}$) grown on a highly conductive n-type substrate. In contrast, the epi-stack of the PiN diode differs only by the absence of the gain layer; the other layers are identical to those in the LGAD (see \figurename~\ref{fig:PIN_3D_Cross_Section}). \figurename~\ref{fig:W2_SIMS} shows the doping concentrations of aluminum and nitrogen measured using secondary ion mass spectrometry (SIMS). The measured doping concentrations are consistent with the design values. \figurename~\ref{fig:ElectricField} shows the estimated electric field distribution in the LGAD and PiN diode by TCAD simulation. The peak electric field in the gain layer of the LGAD is about 2.8 MV/cm at the 600~V reverse bias voltage, and this electric field is sufficient to cause low-gain multiplication of carriers.

Circular mesas varying from 75 to 600 $\mu$m in diameter were formed by an area-selective etch of the 4H-SiC stack to the drift region using a Reactive Ion Etcher (RIE). The samples are prepared by a standard solvent cleaning procedure with subsequent RCA cleaning before being thermally oxidized at 1100$^{\circ}$C to form a SiO$_2$ passivation layer. Approximately 500 nm of PECVD SiO$_2$ is deposited at 300$^{\circ}$C to increase the thickness of the passivation layer. The oxide is then selectively etched to expose the top of the 4H-SiC mesas. A Ti/Al (50/75 nm) metal stack is deposited through e-beam evaporation on the epi-side, followed by a deposition of Ti/Ni (10/100 nm) on the substrate side. The devices are annealed at 1000$^{\circ}$C for 5 minutes in a N$_2$ environment to create low-resistivity contacts. Finally, Al (300 nm) and a Ti/Ni/Au (30/50/50 nm) metal stack are deposited as probe metal using e-beam evaporation on the epi- and substrate side respectively.

\section{I-V and C-V}

\begin{figure*}[htb]
    \centering
    \subfigure[]{ \label{fig:IV_adj}
        \includegraphics[scale=0.5]{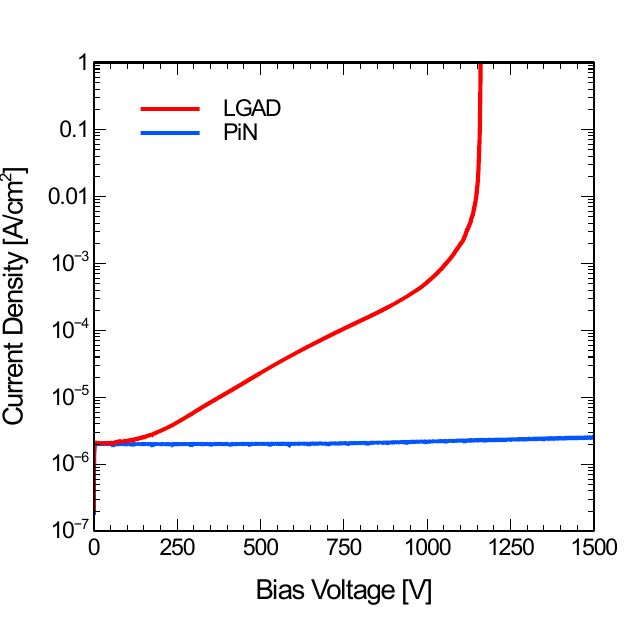}}  
    \subfigure[]{ \label{fig:CV_adj}
        \includegraphics[scale=0.5]{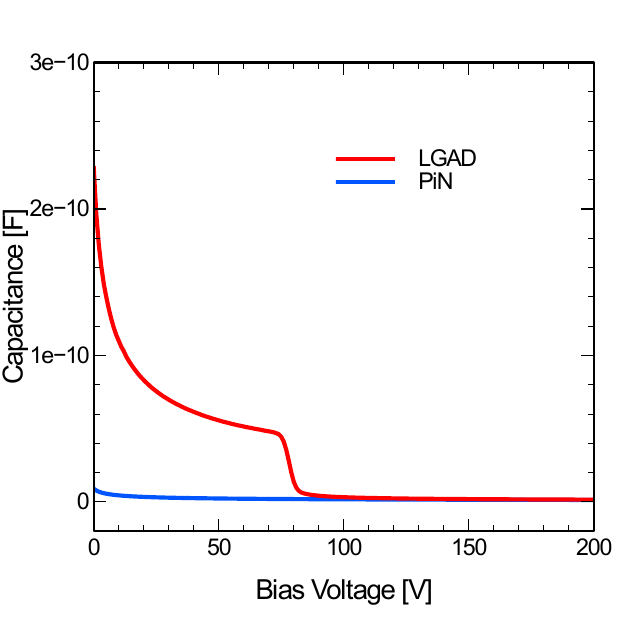}}
    \subfigure[]{ \label{fig:Inv_CV_adj}
        \includegraphics[scale=0.5]{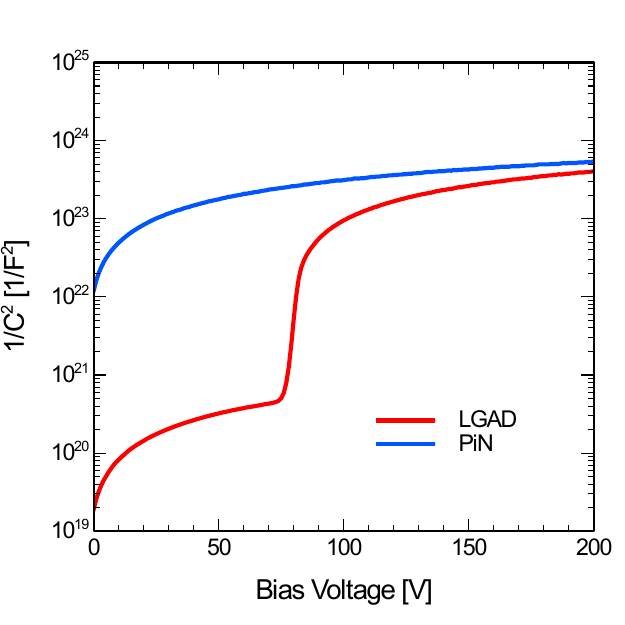}}
    \caption{(a) I-V characteristics of a 75 $\mu$m pad diameter 4H-SiC LGAD and PiN diode under reverse bias. Devices were measured in the dark. (b) C-V and (c) inverse C-V characteristics of a 600 $\mu$m pad diameter 4H-SiC LGAD and PiN diode under reverse bias.}
\end{figure*}

I-V characteristics of the fabricated 4H-SiC LGAD and PiN diode were measured using a Keithley 2657A high power source measure unit (SMU). Measurements were conducted while covering the device under test (DUT) with Fluorinert FC-70 to prevent arcing and to accurately measure the breakdown voltage of the DUT. The LGAD showed a higher leakage current compared to the PiN diode, as shown in \figurename~\ref{fig:IV_adj}. This supports the presence of a functional gain layer as the gain layer is expected to multiply any carriers contributing to the dark current. The breakdown voltage was measured to be 1160 V for a 75 $\mu$m diameter 4H-SiC LGAD. Future designs will incorporate additional edge termination features to increase the breakdown voltage of the LGAD.

C-V characteristics were measured using a Keithley 4200 semiconductor parameter analyzer. Compared to the 4H-SiC PiN diode, the 4H-SiC LGAD displayed a characteristic "step" feature at 77 V seen in \figurename~\ref{fig:CV_adj}, which is in close agreement with the simulated results and confirms the presence of the gain layer in the 4H-SiC LGAD. A gain layer doping level of 2.84$\times$10$^{17}$ cm$^{-3}$ and drift layer doping levels of 3.78$\times$10$^{14}$ cm$^{-3}$ and 4.86$\times$10$^{14}$ cm$^{-3}$ for the 600 $\mu$m diameter 4H-SiC LGAD and PiN diode, respectively, were extracted from linear fits of the inverse C-V data. The net doping levels $n_{net}$ are calculated using the slope of the linear fits and the equation

\begin{equation}
    n_{net} = \frac{2}{q\varepsilon_s\varepsilon_0 A^2 \left(\frac{d(1/C^2)}{dV}\right)}
    \label{eqn:doping}
\end{equation}

\noindent where $q$ = 1.602$\times$10$^{-19}$ C, $\varepsilon_s$ = 10.32, $\varepsilon_0$ = 8.85$\times$10$^{-14}$ F/cm, $A$ is the area of the device, $C$ is the capacitance, and $V$ is the applied voltage. The calculated net doping level values for the gain layer is close to the target value of 3$\times$10$^{17}$ cm$^{-3}$ and agrees with the SIMS results shown in \figurename~\ref{fig:W2_SIMS}. The deviations of the drift region doping levels recorded in SIMS from the calculated values are due to the detection limit for the SIMS measurements for N and Al that are 2$\times$10$^{15}$ cm$^{-3}$ and 3$\times$10$^{14}$ cm$^{-3}$ respectively. The electrical characteristics of both the 4H-SiC LGAD and PiN diode not only display proper behavior but also confirm the device structures shown in \figurename~\ref{fig:LGAD_3D_Cross_Section} and \figurename~\ref{fig:PIN_3D_Cross_Section}.

\section{Radiation Response Testing} \label{sec:rad}

%%\subsubsection{Test setup}

\begin{figure}[htb]
    \centering
    \includegraphics[scale=0.32]{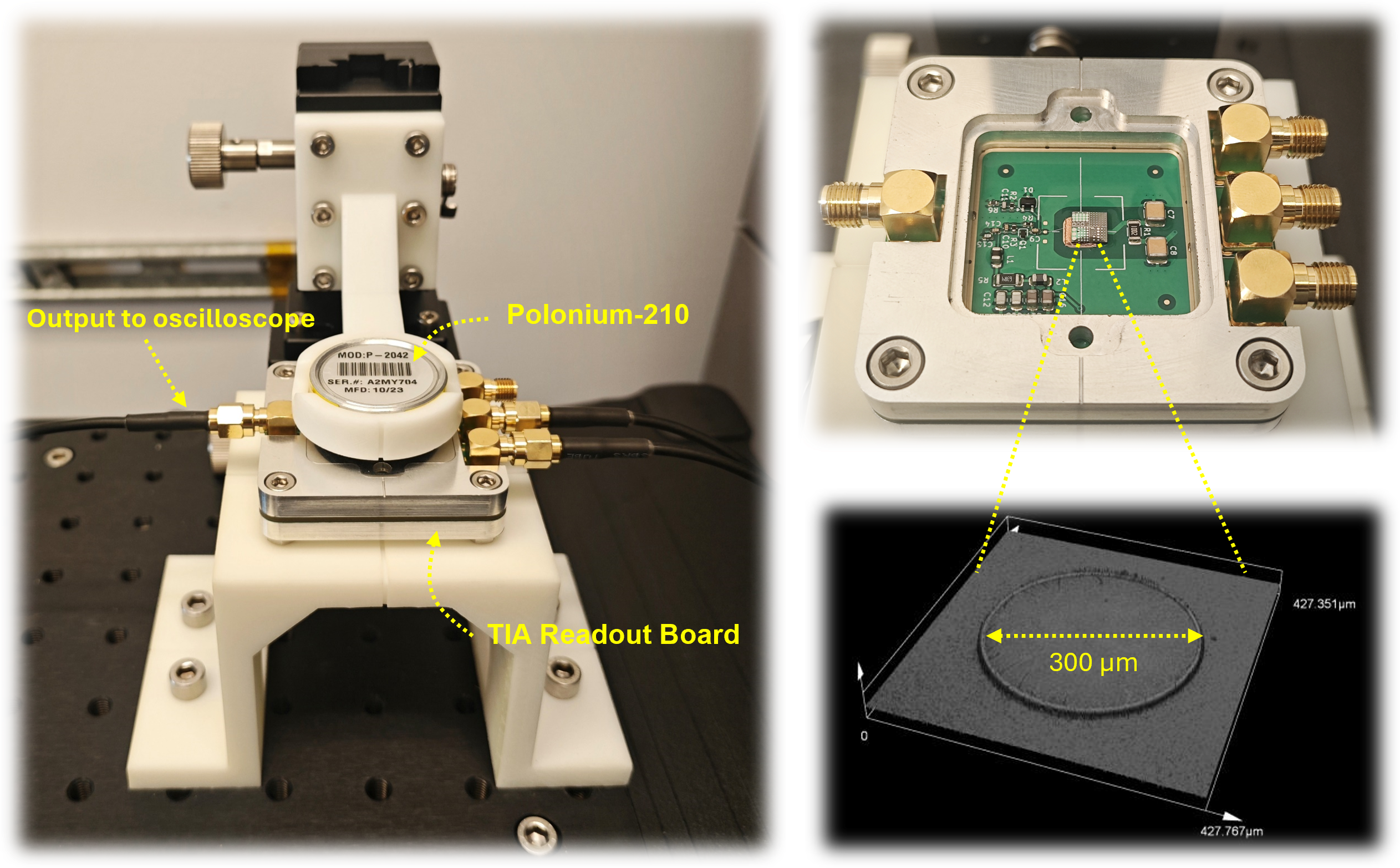}
    \caption{Test setup for the $\alpha$-particle source ($^{210}_{84}\rm{Po}$).}
    \label{fig:Alpha_Setup}
\end{figure}

To demonstrate the charge gain properties of the 4H-SiC LGAD, we used an $\alpha$-particle source and a custom readout board with a transimpedance amplifier (TIA) to measure the signal response of both the 4H-SiC LGAD and the 4H-SiC PiN diode with the same size. The backside of the DUT is attached to the readout board using copper foil conductive tape. Subsequently, the top pad (anode) of the device is connected to the TIA circuit via wire bonding. The bias voltage is applied to the bottom pad (cathode) of the device through the bias pad of the readout board (see \figurename~\ref{fig:Alpha_Setup}). The signal generated by the incident $\alpha$ particles is amplified by the TIA circuit and then output to a high-speed oscilloscope with a 2.5 GHz bandwidth and a sampling rate of 20 GSa/s.

\begin{figure}[htb]
    \centering
    \includegraphics[scale=0.41]{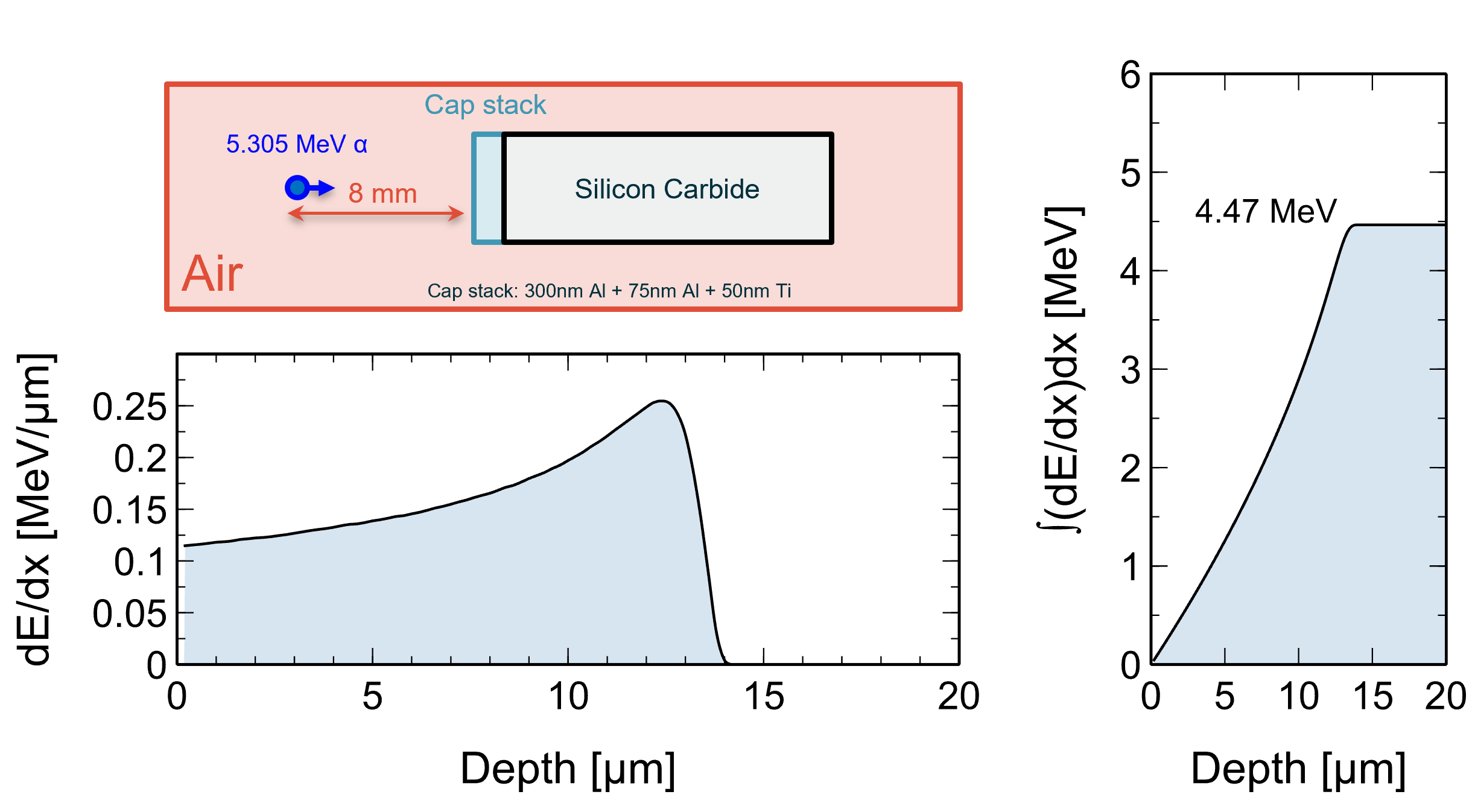}
    \caption{Energy deposition distribution of $\alpha$ particles in silicon carbide by SRIM simulation.}
    \label{fig:SRIM_Sim}
\end{figure}

The $\alpha$-radiation source used in this work is $^{210}_{84}\rm{Po}$, which decays into $^{206}_{82}\rm{Pb}$ (99.999\%) while releasing $\alpha$ particles with an energy of 5.305 MeV. Figure \ref{fig:SRIM_Sim} shows the energy deposition distribution of $\alpha$ particles in SiC, simulated using SRIM \cite{SRIM}. The energy of the $\alpha$ particles degrades due to interactions from atoms in the air between the source and sample as well as the metal stack deposited on the devices. Ultimately, the energy deposited in the SiC is 4.47 MeV according to the SRIM simulation. The simulation results also indicate that the penetration depth of the 4.47 MeV $\alpha$ particles in SiC is approximately 14 µm. Therefore, as long as an appropriate voltage is applied to deplete the device to a depth exceeding 14 µm, all electron-hole pairs generated by the energy deposited from the $\alpha$ particles can theoretically be collected. This conclusion will be validated in the subsequent measurement results.

% \subsubsection{Charges collection for $\alpha$ particles}

\begin{figure*}[htb]
    \centering
    \subfigure[]{ \label{fig:20240712_PIN_G2_W5_5O_U1_SU34_300UM_52_Alpha_600V_waveform_heatmap}
        \includegraphics[scale=0.45]{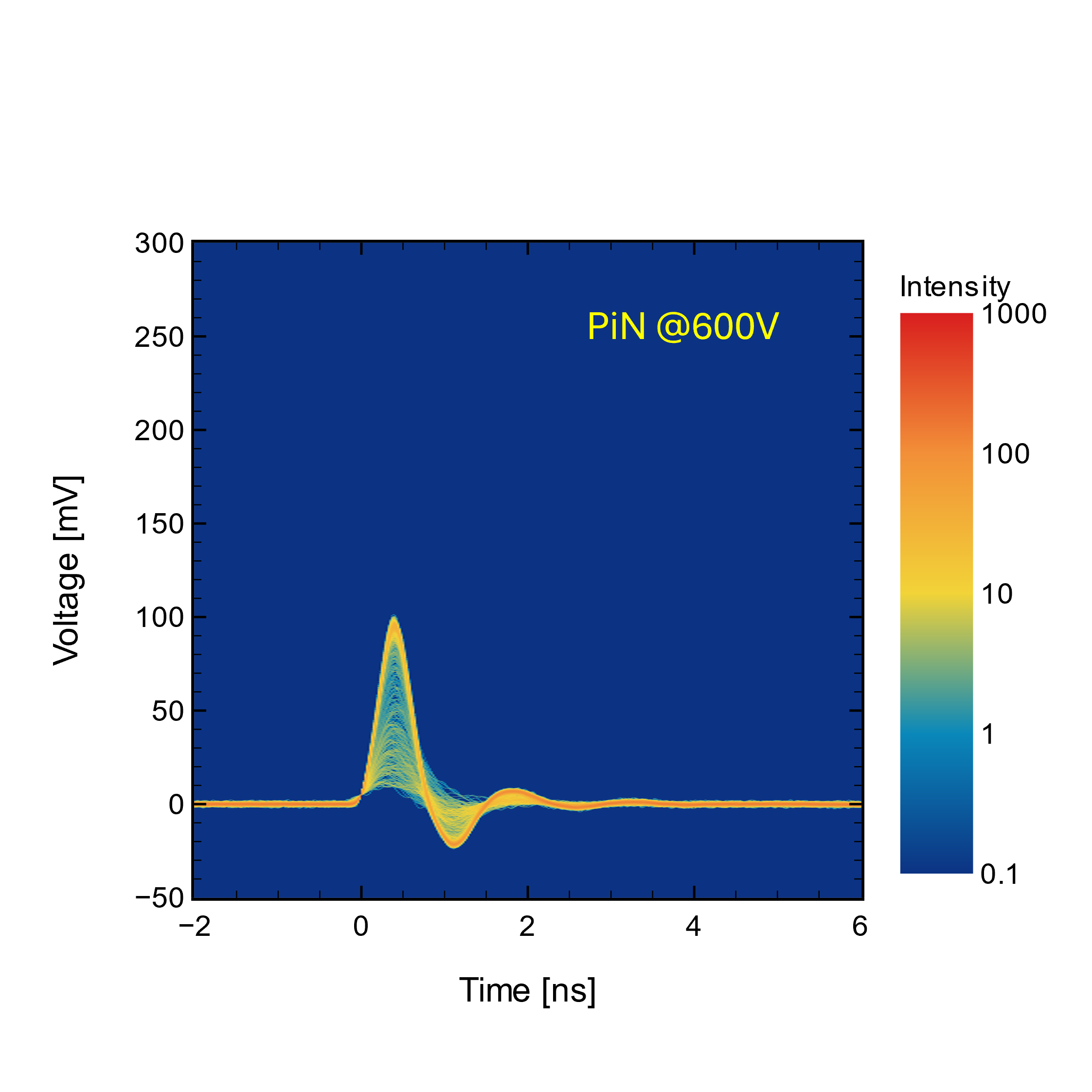}}  
    \subfigure[]{ \label{fig:20240712_LGAD_G2_W2_2I_U2_SU34_300UM_22_Alpha_600V_waveform_heatmap}
        \includegraphics[scale=0.45]{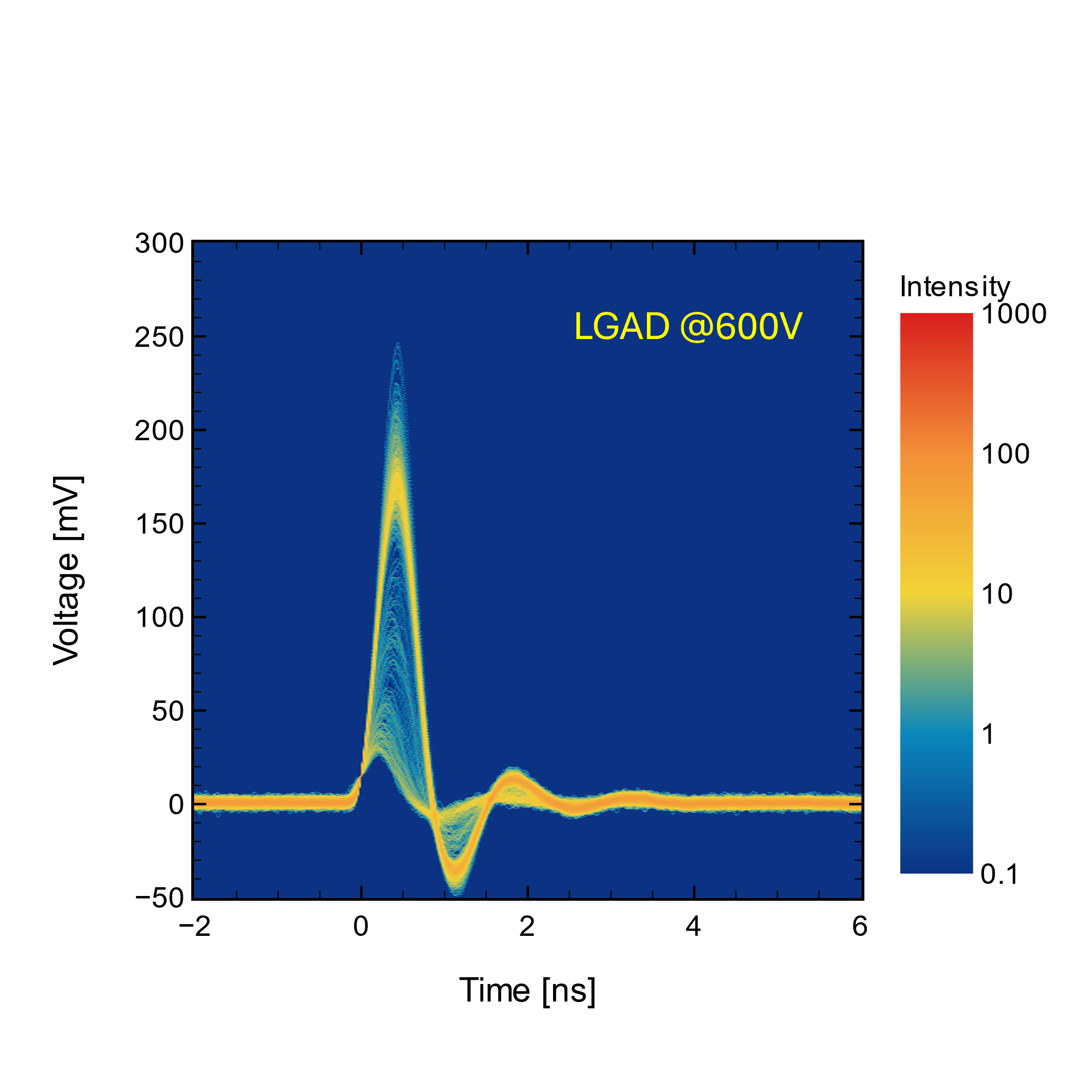}}
    \caption{Signal pulses intensity distribution of $\alpha$ particles: (a)~4H-SiC PiN diode at 600~V and (b)~4H-SiC LGAD at 600~V.}
\end{figure*}

\begin{figure*}[htb]
    \centering
    \subfigure[]{ \label{fig:20240712_pin_lgad_integral_fit_gauss}
        \includegraphics[scale=0.45]{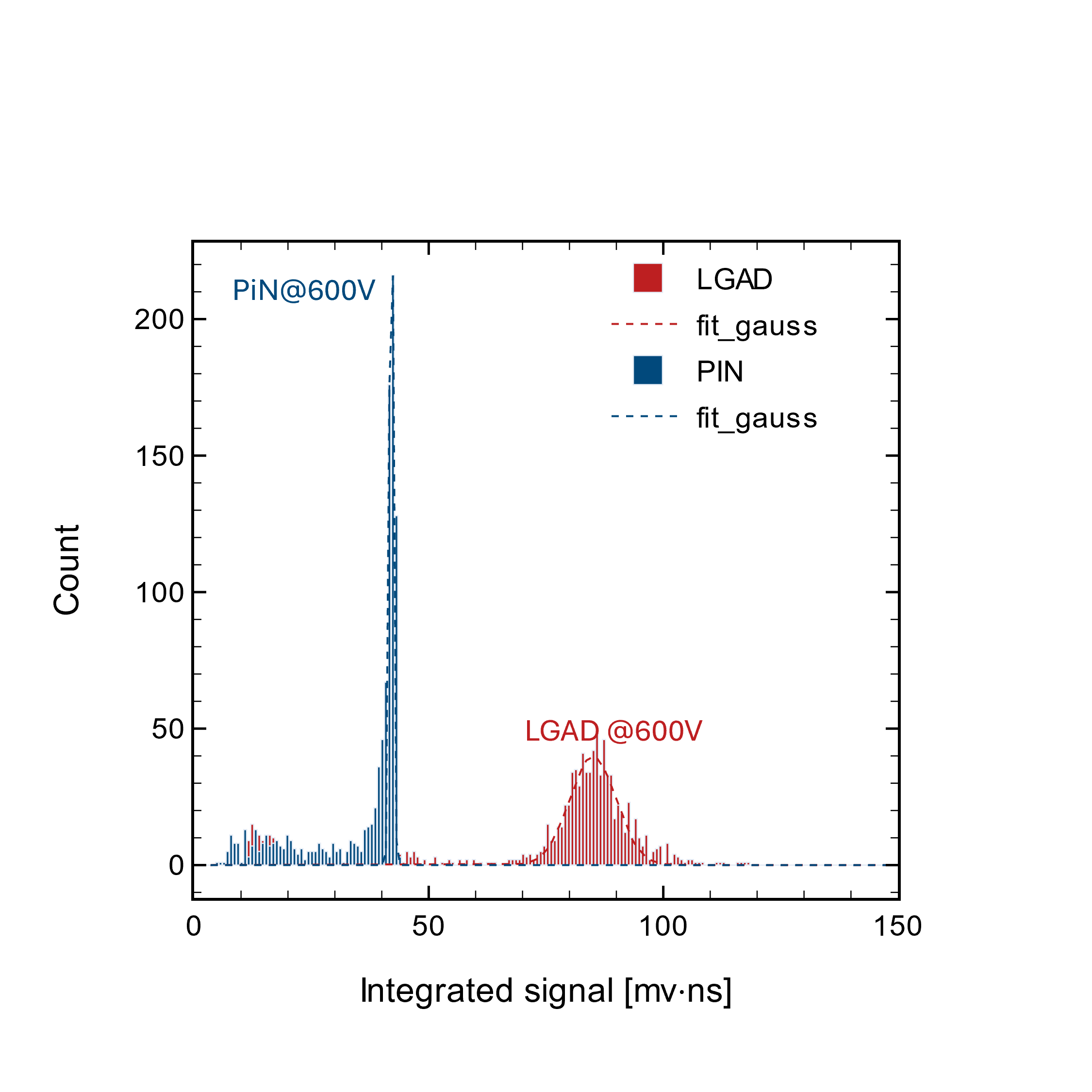}}  
    \subfigure[]{ \label{fig:20240712_pin_lgad_signal_integral}
        \includegraphics[scale=0.45]{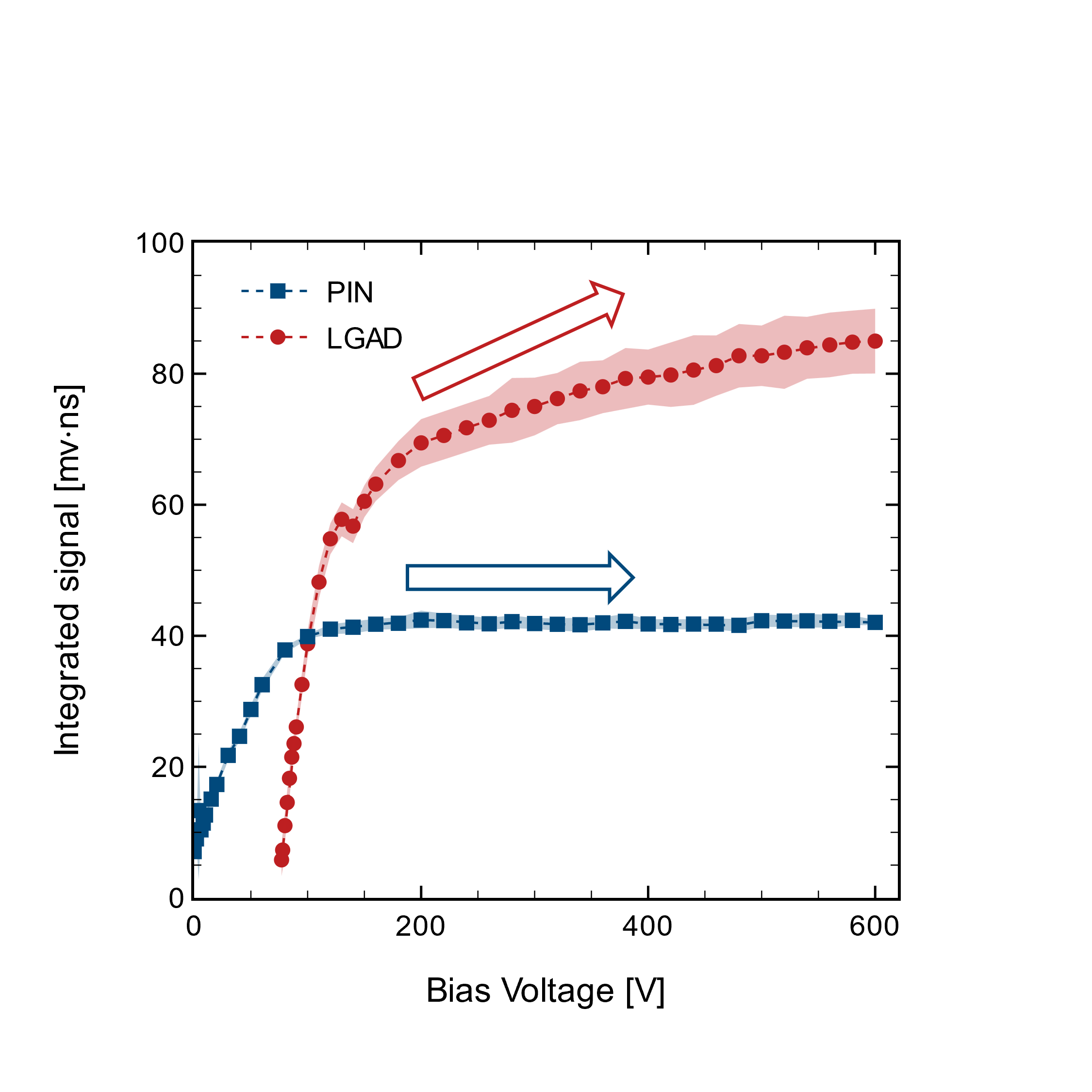}}
    \caption{(a)~The distribution of integrated signal of the 4H-SiC PiN diode (blue) and 4H-SiC LGAD (red) fitted by Gaussian function in a local range with the 600V bias voltage. (b)~integrated signal vs bias voltage of the 4H-SiC PiN diode (blue) and 4H-SiC LGAD (red).}
\end{figure*}

\figurename~\ref{fig:20240712_PIN_G2_W5_5O_U1_SU34_300UM_52_Alpha_600V_waveform_heatmap} and \figurename~\ref{fig:20240712_LGAD_G2_W2_2I_U2_SU34_300UM_22_Alpha_600V_waveform_heatmap} show the signals of $\alpha$ particles acquired by the oscilloscope. \figurename~\ref{fig:20240712_PIN_G2_W5_5O_U1_SU34_300UM_52_Alpha_600V_waveform_heatmap} illustrates the signal intensity distribution of the 4H-SiC PiN diode at an applied bias voltage of 600V. \figurename~\ref{fig:20240712_LGAD_G2_W2_2I_U2_SU34_300UM_22_Alpha_600V_waveform_heatmap} shows the signal intensity distribution of the 4H-SiC LGAD also at an applied bias voltage of 600V. Both the PiN diode and LGAD signals have a pulse width of 1 ns and a rise time about 300~ps from 10\%-90\% of amplitude. However, the signal amplitude of the LGAD is higher than that of the PiN diode. According to the Shockley-Ramo theorem \cite{Shockley_Ramo}, the transient current is proportional to the charge quantity. Therefore, comparing the signal amplitudes of the PiN diode and LGAD confirms that there is multiplication of charge carriers generated in the LGAD.

The signal pulse captured by the oscilloscope is the amplified output voltage resulting from the original current pulse signal passing through the TIA circuit. This relationship can be described by $V_{out} = I_{in} \times Z$, where $Z$ is the impedance of the amplifier. Therefore, the collected charge $Q$ is proportional to the integral of the output voltage pulse, where $ Q = \int I_{in}~dt \propto \int V_{out}~dt$. \figurename~\ref{fig:20240712_pin_lgad_integral_fit_gauss} shows the distribution of integrated voltage pulses from \figurename~\ref{fig:20240712_LGAD_G2_W2_2I_U2_SU34_300UM_22_Alpha_600V_waveform_heatmap} and \figurename~\ref{fig:20240712_LGAD_G2_W2_2I_U2_SU34_300UM_22_Alpha_600V_waveform_heatmap}. Integrating the signals shows that the amount of charge collected from $\alpha$-particle ionization in the LGAD is higher than that generated in PiN diode. This demonstrates that low-gain multiplication of charge carriers occurred in the LGAD. Additionally, it can be observed that the charge collection spectrum of the LGAD is broader compared to that of PiN diode. This may be due to homogeneity issues in the gain layer, causing different gain values at different $\alpha$ particle incident positions and the fluctuations in the multiplication process.  We intent to verify this hypothesis  in the future using ultraviolet transient current technique (UV-TCT) \cite{SiC_TCT_Report}.

\figurename~\ref{fig:20240712_pin_lgad_signal_integral} shows the curves of the integrated signals in the LGAD and PiN diode as a function of the bias voltage. Since the integrated signal is proportional to the collected charge, the integrated signal curves of the PiN diode and LGAD correspond to the variation of collected charge with voltage. As shown in \figurename~\ref{fig:20240712_pin_lgad_signal_integral}, when the bias voltage increases to approximately 100V, the collected charge in the PiN diode no longer increases with the voltage. This is because the depletion depth in the PiN diode exceeds 14 µm at this voltage and is greater than the penetration depth of $\alpha$ particles in SiC (\figurename~\ref{fig:SRIM_Sim}). Therefore, it can be approximated that the charge generated by $\alpha$-particle ionization is fully collected. In contrast to the PiN diode, the LGAD has a 0.5 µm highly doped gain layer, which needs to be depleted first. Therefore, at low voltages and with a thin depletion layer, the $\alpha$-particle signal is minimal in the LGAD. As the applied bias voltage continues to increase and exceeds the depletion voltage of the gain layer, the collected charge begins to increase rapidly. Additionally, the electric field in the gain layer causes charge carriers to undergo multiplication, and since the multiplication field increases with the bias voltage, the gain also increases. Consequently, the collected charge in the LGAD is observed to be greater than that in the PiN diode and continues to increase with increasing voltage. In summary, we have successfully achieved low-gain charge carrier multiplication in the 4H-SiC LGAD.

However, the charge collection measurements indicate that the gain of the 4H-SiC LGAD is not very high, approximately between 2-3. The ideal gain for LGADs should be between 10 and 100. One possible reason for this could be the low doping concentration of the gain layer. Another potential reason could be the space charge effect caused by the large amount of charge with high density generated by $\alpha$-particle ionization, that could suppress the gain. The space charge effect has also been observed in Si LGADs \cite{Gain_Reduction}, where $\alpha$ particles exhibit lower gain \cite{Diff_Source_Test} than $\beta$ particles, $\gamma$-rays and X-rays. This will be further investigated in the future using UV-TCT and by comparing the charge collection results of 4H-SiC LGADs with higher gain layer doping concentrations.

\section{Conclusion}
In conclusion, an operational 4H-SiC LGAD with high breakdown voltage, low leakage current, and low-gain carrier multiplication has been achieved. Having such a device structure also presents an opportunity to refine models of avalanche multiplication in 4H-SiC. Further studies are planned on the use of various edge termination structures and their impacts on device performance. The fabrication of multi-pixel designs (i.e., AC-LGADs, TI-LGADs, i-LGADs) and future radiation reliability characterization are also planned. As 4H-SiC fabrication technology matures, 4H-SiC could play a greater role in LGAD technology and in solid-state detectors for future high-energy and nuclear physics applications.

\section*{CRediT authorship contribution statement}
\textbf{Tao Yang}: Writing – original draft, Methodology, Investigation, Formal analysis, Data curation, Conceptualization. \textbf{Ben Sekely}: Writing – original draft, Methodology, Investigation, Formal analysis, Data curation. \textbf{Yashas Satapathy}: Writing – original draft, Methodology, Investigation, Formal analysis, Data curation. \textbf{Greg Allion}: Writing - review \& editing, Methodology, Investigation, Formal analysis. \textbf{Philip Barletta}: Writing - review \& editing, Methodology, Investigation, Formal analysis. \textbf{Carl Haber}: Writing – review \& editing, Supervision, Resources, Project administration, Methodology, Funding acquisition, Formal analysis, Conceptualization. \textbf{Steve Holland}: Writing - review \& editing, Methodology, Investigation, Formal analysis. \textbf{John F. Muth}: Writing – review \& editing, Supervision, Resources, Project administration, Methodology, Funding acquisition, Formal analysis, Conceptualization. \textbf{Spyridon Pavlidis}: Writing - review \& editing, Methodology, Investigation, Formal analysis. \textbf{Stefania Stucci}: Writing - review \& editing, Methodology, Investigation, Formal analysis. 

\section*{Declaration of competing interest}
The authors declare that they have no known competing financial interests or personal relationships that could have appeared to influence the work reported in this paper.

\section*{Acknowledgment}
This material is based upon work supported by the U.S. Department of Energy, Office of Science, Office of High Energy Physics, under Contract Number DE-AC02-05CH11231 and Award Number DE-SC0024252. Additional support was provided by the National Science Foundation under contract No. ECCS-1542015. We are grateful to Maurice Garcia-Sciveres, Timon Heim, Peter Sorensen, Azriel Goldschmidt and Thorsten Stezelberger of LBNL for providing experimental equipment. We also thank James Boldi and Phathakone Sanethavong of LBNL for their help in machining and wire bonding. We would like to thank the NCSU Nanofabrication Facility for their knowledge and expertise used in fabrication processing.

\bibliographystyle{unsrt}
\bibliography{p1_sic_lgad}

\begin{thebibliography}{10}

\bibitem{CNM_LGAD}
G.~Pellegrini et~al.
\newblock {Technology developments and first measurements of Low Gain Avalanche Detectors (LGAD) for high energy physics applications}.
\newblock {\em Nucl. Instrum. Methods A}, 765, 2014.
\newblock doi: \href{https://doi.org/10.1016/j.nima.2014.06.008}{10.1016/j.nima.2014.06.008}.

\bibitem{HPK_LGAD}
H.~F.~W. Sadrozinski et~al.
\newblock {4D tracking with ultra-fast silicon detectors}.
\newblock {\em Rep. Prog. Phys.}, 81(2), 2018.
\newblock doi: \href{https://iopscience.iop.org/article/10.1088/1361-6633/aa94d3}{10.1088/1361-6633/aa94d3}.

\bibitem{FBK_LGAD}
V.~Sola et~al.
\newblock {First FBK production of 50 um ultra-fast silicon detectors}.
\newblock {\em Nucl. Instrum. Methods A}, 924, 2019.
\newblock doi: \href{https://doi.org/10.1016/j.nima.2018.07.060}{10.1016/j.nima.2018.07.060}.

\bibitem{BNL_LGAD}
G.~Giacomini et~al.
\newblock {Development of a technology for the fabrication of Low-Gain Avalanche Diodes at BNL}.
\newblock {\em Nucl. Instrum. Methods A}, 934, 2019.
\newblock doi: \href{https://doi.org/10.1016/j.nima.2019.04.073}{10.1016/j.nima.2019.04.073}.

\bibitem{KeweiWU_LGAD}
K.~Wu et~al.
\newblock {Design of Low Gain Avalanche Detectors (LGAD) with 400 keV ion implantation energy for multiplication layer fabrication}.
\newblock {\em Nucl. Instrum. Methods A}, 984, 2020.
\newblock doi:\href{https://doi.org/10.1016/j.nima.2020.164558}{10.1016/j.nima.2020.164558}.

\bibitem{YunyunFAN_NDL}
Y.Y. Fan et~al.
\newblock {Radiation hardness of the low gain avalanche diodes developed by NDL and IHEP in China}.
\newblock {\em Nucl. Instrum. Methods A}, 984, 2020.
\newblock doi:\href{https://doi.org/10.1016/j.nima.2020.164608}{10.1016/j.nima.2020.164608}.

\bibitem{ATLAS_HGTD}
{ATLAS Collaboration}.
\newblock {Technical Proposal: A High-Granularity Timing Detector for the ATLAS Phase-II Upgrade}.
\newblock Technical report, 2018.
\newblock \href{https://cds.cern.ch/record/2623663}{CERN-LHCC-2018-023. LHCC-P-012}.

\bibitem{CMS_ETL}
{CMS Collaboration}.
\newblock {A MIP Timing Detector for the CMS Phase-2 Upgrade}.
\newblock Technical report, 2019.
\newblock \href{https://cds.cern.ch/record/2667167}{CERN-LHCC-2019-003. CMS-TDR-020}.

\bibitem{EIC_Concept}
{EIC Collaboration}.
\newblock {Electron Ion Collider Conceptual Design Report}.
\newblock Technical report, 2019.
\newblock url:\href{https://www.bnl.gov/ec/files/eic_cdr_final.pdf}{https://www.bnl.gov/ec/}.

\bibitem{LGAD_RAD_I_G}
G.~Kramberger et~al.
\newblock {Radiation effects in Low Gain Avalanche Detectors after hadronirradiations}.
\newblock {\em Journal of Instrumentation}, 10(07), 2015.
\newblock doi: \href{https://doi.org/10.1088/1748-0221/10/07/p07006}{10.1088/1748-0221/10/07/p07006}.

\bibitem{Ar_Model}
M.~Ferrero et~al.
\newblock {Radiation resistant LGAD design}.
\newblock {\em {Nucl. Instrum. Methods A}}, 919, 2019.
\newblock doi:\href{https://doi.org/10.1016/j.nima.2018.11.121}{10.1016/j.nima.2018.11.121}.

\bibitem{SiC_Rad}
Joan~Marc Rafí et~al.
\newblock {Electron, Neutron, and Proton Irradiation Effects on SiC Radiation Detectors}.
\newblock {\em IEEE Transactions on Nuclear Science}, 67, 2020.
\newblock {doi:\href{https://ieeexplore.ieee.org/document/9217477}{10.1109/TNS.2020.3029730}}.

\bibitem{SiC_Alpha_Time}
Xiaodong Zhang et~al.
\newblock {Characterizing the Timing Performance of a Fast 4H-SiC Detector With an $^{241}$ Am Source}.
\newblock {\em IEEE Transactions on Nuclear Science}, 60, 2013.
\newblock doi:\href{https://doi.org/10.1109/TNS.2013.2260652}{10.1109/TNS.2013.2260652}.

\bibitem{Tao_NJU_PIN_Time}
T.Yang et~al.
\newblock {Time Resolution of the 4H-SiC PIN Detector}.
\newblock {\em Frontiers in Physics}, 10, 2022.
\newblock {doi:\href{https://www.frontiersin.org/article/10.3389/fphy.2022.718071}{10.3389/fphy.2022.718071}}.

\bibitem{First_SiC_LGAD_Report}
T.Yang et~al.
\newblock {Time resolution of 4H-SiC PIN and simulation of 4H-SiC LGAD}, 2021.
\newblock { The 38th RD50 Workshop, url:\href{https://indico.cern.ch/event/1029124/contributions/4411189/}{https://indico.cern.ch/event/1029124/contributions/4411189/}}.

\bibitem{NJU_SiC_LGAD_Report_2}
T.Yang et~al.
\newblock {Development of 4H-SiC Low Gain Avalanche Diode}, 2022.
\newblock { The 40th RD50 Workshop, url:\href{https://indico.cern.ch/event/1157463/contributions/4922751/}{https://indico.cern.ch/event/1157463/contributions/4922751/}}.

\bibitem{IHEP_SiC_LGAD_Report}
C.Wang et~al.
\newblock {Development of 4H-SiC Low-Gain Avalanche Detector}, 2023.
\newblock { The 42nd RD50 Workshop, url:\href{https://indico.cern.ch/event/1270076/contributions/5450184/}{https://indico.cern.ch/event/1270076/contributions/5450184//}}.

\bibitem{Tao_Design_SiC_LGAD}
T.Yang et~al.
\newblock {Design and simulation of {4H}-{SiC} low gain avalanche diode}.
\newblock {\em Nucl. Instrum. Methods A}, 1056, 2023.
\newblock doi:\href{https://doi.org/10.1016/j.nima.2023.168677}{10.1016/j.nima.2023.168677}.

\bibitem{SRIM}
James~F. Ziegler et~al.
\newblock {SRIM – The stopping and range of ions in matter (2010)}.
\newblock {\em Nucl. Instrum. Methods B}, 268, 2010.
\newblock {doi:\href{https://doi.org/10.1016/j.nimb.2010.02.091}{10.1016/j.nimb.2010.02.091}}.

\bibitem{Shockley_Ramo}
W.~Shockley.
\newblock {{Currents to Conductors Induced by a Moving Point Charge}}.
\newblock {\em Journal of Applied Physics}, 9, 1938.
\newblock doi:\href{https://doi.org/10.1063/1.1710367}{10.1063/1.1710367}.

\bibitem{SiC_TCT_Report}
Ivan~Lopez Paz et~al.
\newblock {TCT Study on the effect of epitaxial graphene contacts in SiC detectors}, 2022.
\newblock { The 41st RD50 Workshop, url:\href{https://indico.cern.ch/event/1132520/contributions/5148174/}{https://indico.cern.ch/event/1132520/contributions/5148174/}}.

\bibitem{Gain_Reduction}
E.~Currás et~al.
\newblock {Gain reduction mechanism observed in Low Gain Avalanche Diodes}.
\newblock {\em Nucl. Instrum. Methods A}, 1031, 2022.
\newblock {doi:\href{https://doi.org/10.1016/j.nima.2022.166530}{j.nima.2022.166530}}.

\bibitem{Diff_Source_Test}
Gabriele Giacomini et~al.
\newblock {Spectroscopic performance of Low-Gain Avalanche Diodes for different types of radiation}.
\newblock {\em Nucl. Instrum. Methods A}, 1066, 2024.
\newblock {doi:\href{https://doi.org/10.1016/j.nima.2024.169605}{j.nima.2024.169605}}.

\end{thebibliography}

\end{document}